\documentclass[aps,prb,twocolumn,superscriptaddress]{revtex4-1}

\usepackage{amsmath}
\usepackage{mathptmx}           
\usepackage{graphicx}          
\usepackage{tikz}               
\usetikzlibrary{plotmarks}
\usepackage{array}              
\usepackage[T1]{fontenc}
\usepackage{url}                
\usepackage{bbold}
\usepackage{enumitem}

\begin{document}

\title{The effect of magnetism and temperature on the stability of \(\text{(}\text{Cr}_{x}\text{,}\text{V}_{{1-x}}\text{)}_{\text{2}}\text{AlC}\) phases}

\author{Jo\'as Grossi}
\affiliation{Facultad de Ciencias Exactas y Naturales, Universidad Nacional de Cuyo, Mendoza, 5500 Argentina}
\affiliation{Theory of Condensed Matter,
             Cavendish Laboratory, University of Cambridge, 
             J. J. Thomson Ave, Cambridge CB3 0HE, UK}

\author{Shafqat H. Shah}
\affiliation{Department of Materials Science and Metallurgy, University of Cambridge, 27 Charles Babbage Rd, Cambridge CB3 0FS, UK} 

\author{Emilio Artacho}
\affiliation{Theory of Condensed Matter,
             Cavendish Laboratory, University of Cambridge, 
             J. J. Thomson Ave, Cambridge CB3 0HE, UK}
\affiliation{CIC Nanogune and DIPC, Tolosa Hiribidea 76, 20018 San Sebasti\'an, Spain}
\affiliation{Basque Foundation for Science Ikerbasque, 48013 Bilbao, Spain}

\author{Paul D. Bristowe}
\affiliation{Department of Materials Science and Metallurgy, University of Cambridge, 27 Charles Babbage Rd, Cambridge CB3 0FS, UK} 
\date{\today}

\begin{abstract}
The stability of  $\text{(}\text{Cr}_\text{x}\text{,}\text{V}_{\text{1-x}}\text{)}_{\text{2}}\text{AlC}$ MAX phases, materials of interest for a variety of magnetic as well as high temperature applications, has been studied using density-functional-theory first-principles calculations. The enthalpy of mixing predicts these alloys to be unstable towards unmixing at 0 K. The calculations also predict, however, that these phases would be thermally stabilised by configurational entropy at temperatures well below the values used for synthesis. The temperature $T_s$ below which they become unstable is found to be quite sensitive to the presence of magnetic moments on Cr ions, as well as to the material's magnetic order, in addition to chemical order and composition. Allowing for magnetism, the value of $T_s$  for $\text{(}\text{Cr}_\text{0.5}\text{,}\text{V}_{\text{0.5}}\text{)}_{\text{2}}\text{AlC}$ with chemically disordered Cr and V atoms, is estimated to be between 516 K and 645 K depending on the level of theory, while if constrained to spin-paired, $T_s$ drops to $\sim 142$ K. Antiferromagnetic spin arrangements are found to be favoured at low temperatures, but they are most likely lost at synthesis temperatures, and probably at room temperature as well. However, the combination of antiferromagnetic frustration and configurational disorder should give rise to interesting spin textures at low temperatures.
\end{abstract}
\maketitle

\section{Introduction}

There has been increasing interest in a class of ternary layered carbides and nitrides called MAX phases \cite{Barsoum2013,1Barsoum2000}. Their general chemical formula is $\text{M}_{\text{n+1}}\text{A}\text{X}_{\text{n}}$ where ${n}=1$-3, M is an early transition metal, A is an A-group element and X is most often carbon but sometimes nitrogen. At least 60 examples are known\cite{Eklund2010}. Structurally they are hexagonal and can be described as an array of edge-sharing $\text{MX}_{\text{6}}$ octahedra separated by close-packed layers of A atoms. The spacing between the A layers and the size of the octahedral array is determined by ${n}$. Strong M-X bonds are present and to a lesser extent A-X bonds. The combined effects of crystallography and interatomic bonding result in a number of unique physical and chemical properties such as good machinability, corrosion resistance, high electrical conductivity and tolerance to radiation damage that make MAX phases potential candidates for a wide range of applications\cite{Sun2011,2Barsoum2000,Li2010,Radovic2013,Zhang2006,Wang2009,Lin2007,Zhen2005,Wang2003,Qian2011,Clark2016,Hoffman2012,Allen2012}.

One particular property that has received increasing attention is magnetism mainly due to the possibility of incorporating magnetic elements into the M layers and the prospect of creating multilayer spintronic devices\cite{Ingason2016,Magnuson2017}. Many of the studies performed to date, either theoretical or experimental, have focused on the introduction of transition metal elements which exhibit strong 3d electron correlation effects, such as Fe, Mn or Cr, in an attempt to create spin configurations that exhibit magnetic order\cite{Luo2008,Dahlqvist2015,Mockute2015,Ingason2014,Liu2013}. 

However, the properties and thermodynamic stability of MAX phases can be altered by incorporating further elements into their structure and this is most often done by substitution on the M-site. This may be beneficial from both the phase stability and magnetic point of view. To this end and with magnetism in mind, a number of measurements and calculations have been made on $\text{(}\text{Cr}\text{,}\text{Mn}\text{)}_{\text{2}}\text{XC}$ (X = Al, Ge or Ga) quaternary carbide phases resulting in some cases in the prediction of weak ferromagnetism at low temperatures\cite{Lin2013,Lin2016,Rivin2017,Salikhov2015,Mockute2014,Dahlqvist2011,Liu2014,Ingason2013}. Another possible candidate is the $\text{(}\text{Cr}_\text{x}\text{,}\text{V}_{\text{1-x}}\text{)}_{\text{2}}\text{AlC}$ (0$\leq x \leq$1) solid solution series which has been synthesised\cite{Halim2017,Caspi2015,Schuster1980,Zhou2008} but not completely characterised magnetically. While several studies have been made on the end members of this series including phase stability, magnetism and point defect formation\cite{Shang2014,Dahlqvist2010,Dahlqvist2015,Shah2017,Han2016,Wang2016b}, no systematic investigation of magnetic order has been made over the entire composition range.

The end member that has received most attention is $\text{Cr}_{\text{2}}\text{AlC}$. Recent measurements of its magnetic order have interpreted the spin configuration as very weak ferromagnetic (FM)\cite{Jaouen2013} or canted antiferromagnetic (AFM)\cite{Jaouen2014}. This has not been experimentally resolved yet. 
However, from the theoretical point of view, several groups have searched for possible spin polarised Cr atoms in $\text{Cr}_{\text{2}}\text{AlC}$ using first principles density functional theory (DFT) calculations in attempts to identify its ground state. Considering a small set of FM, AFM, and nonmagnetic (NM) configurations within a single unit cell, a NM solution has been predicted for this material\cite{Sun2003,Schneider2004}. However allowing for unit cell doubling, AFM configurations with antiparallel spins of Cr atoms within the basal plane, were found to be energetically favorable\cite{Dahlqvist2013,2Dahlqvist2015}. In addition, DFT${+U}$ calculations, using the onsite Coulomb repulsion ${U}$ for localized Cr 3d-electrons\cite{Ramzan2011,Dahlqvist2013,2Dahlqvist2015,Wang2016}, were able to stabilise the same magnetic ground states of the AFM configurations and also a FM ground state within a single unit cell. In fact, the AFM configurations were found to be the most stable phases with or without ${+U}$ corrections.
The observed Curie temperature ${T_{\text{C}} \sim 73 \text{K}}$\cite{Jaouen2013} for $\text{Cr}_{\text{2}}\text{AlC}$ would seem to indicate that magnetic effects should not be relevant to the stability of the  $\text{(}\text{Cr}_\text{x}\text{,}\text{V}_{\text{1-x}}\text{)}_{\text{2}}\text{AlC}$ system at temperatures of interest, mainly room or synthesis temperature. We will see below that this is not the case.

The other end member is $\text{V}_{\text{2}}\text{AlC}$ for which a
NM solution has been predicted\cite{Dahlqvist2013,Schneider2006} from first principles. This appears to be consistent with recent measurements which suggest that it is a Pauli paramagnet up to room temperature\cite{Hamm2018}.

\renewcommand{\thetable}{\arabic{table}}
\renewcommand{\tablename}{FIG.}
\begin{table*}[t]
  \begin{tabular}{c c c c c}
${x=1}$ & ${x=0.75}$ & ${x=0.5}$ & ${x=0.25}$ & ${x=0}$ \\ 
\minipage{0.11\textwidth}
  \centering	
  \includegraphics[width=\linewidth]{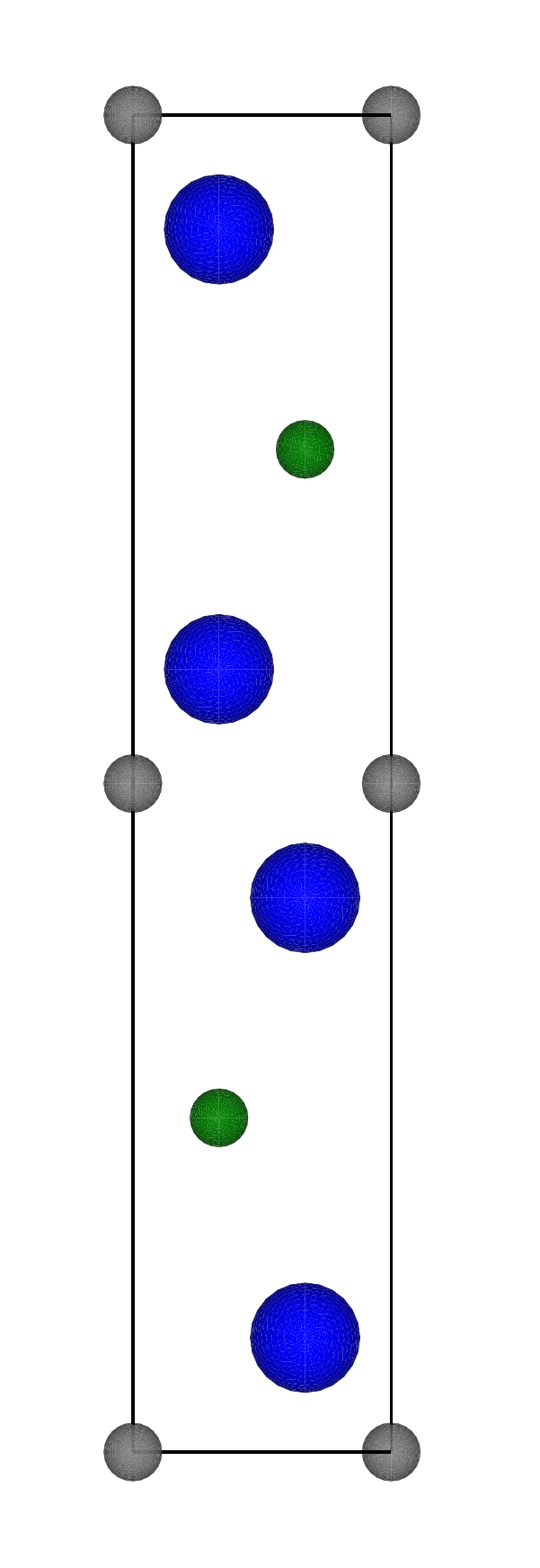}
  Cr${_{2}}$AlC
  \vspace{3mm}
\endminipage
&
\minipage{0.11\textwidth}
  \centering
  \includegraphics[width=\linewidth]{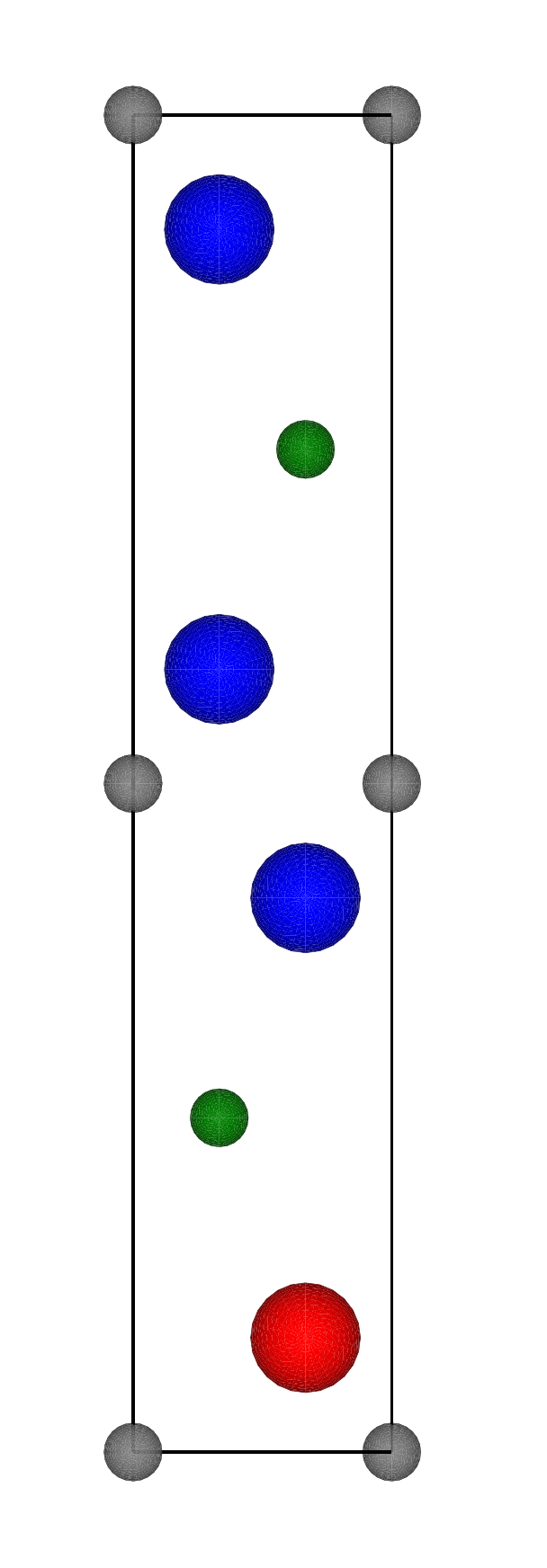}
  CrCrCrV 
  \vspace{3mm}
\endminipage
&
\minipage{0.11\textwidth}
  \centering 
  \includegraphics[height=5.57cm,width=\linewidth]{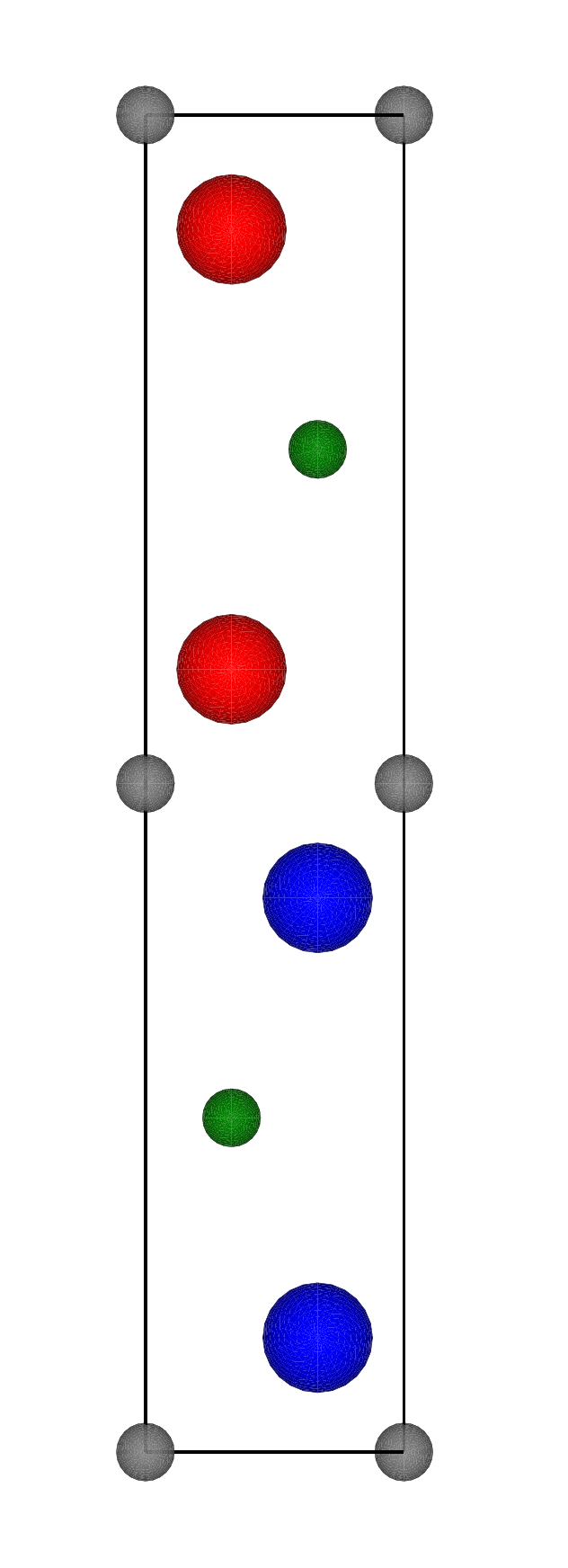}
  VVCrCr
  \vspace{3mm}
\endminipage
\minipage{0.11\textwidth}
  \centering
  \includegraphics[height=5.57cm,width=\linewidth]{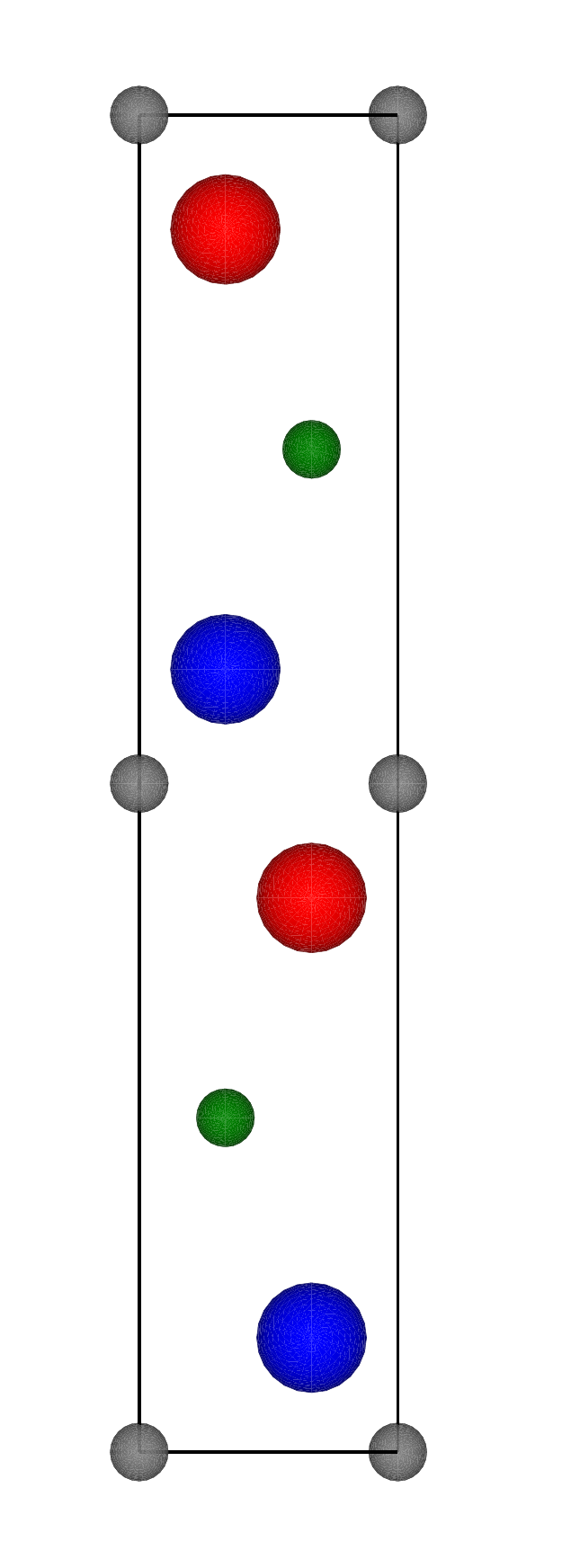}
  VCrVCr
  \vspace{3mm}
\endminipage
\minipage{0.11\textwidth}
  \centering	
  \includegraphics[height=5.57cm,width=\linewidth]{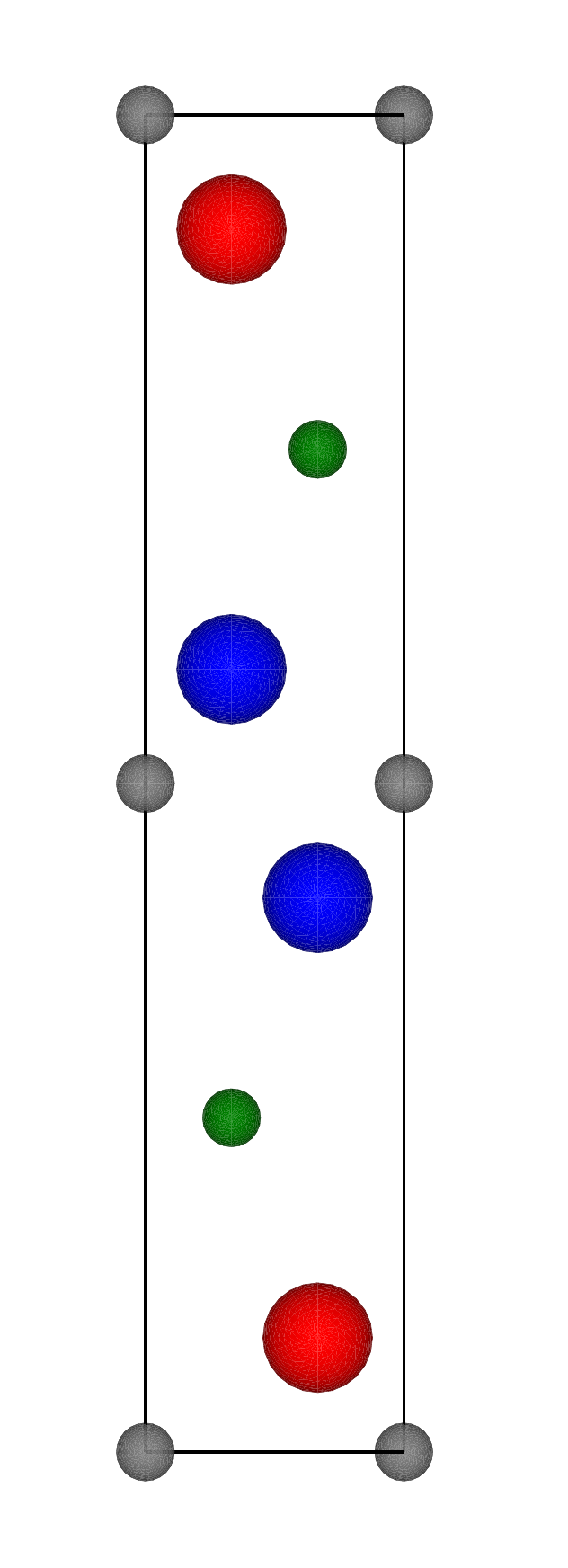}
  VCrCrV 
  \vspace{3mm}
\endminipage
&
\minipage{0.11\textwidth}
  \centering	
  \includegraphics[width=\linewidth]{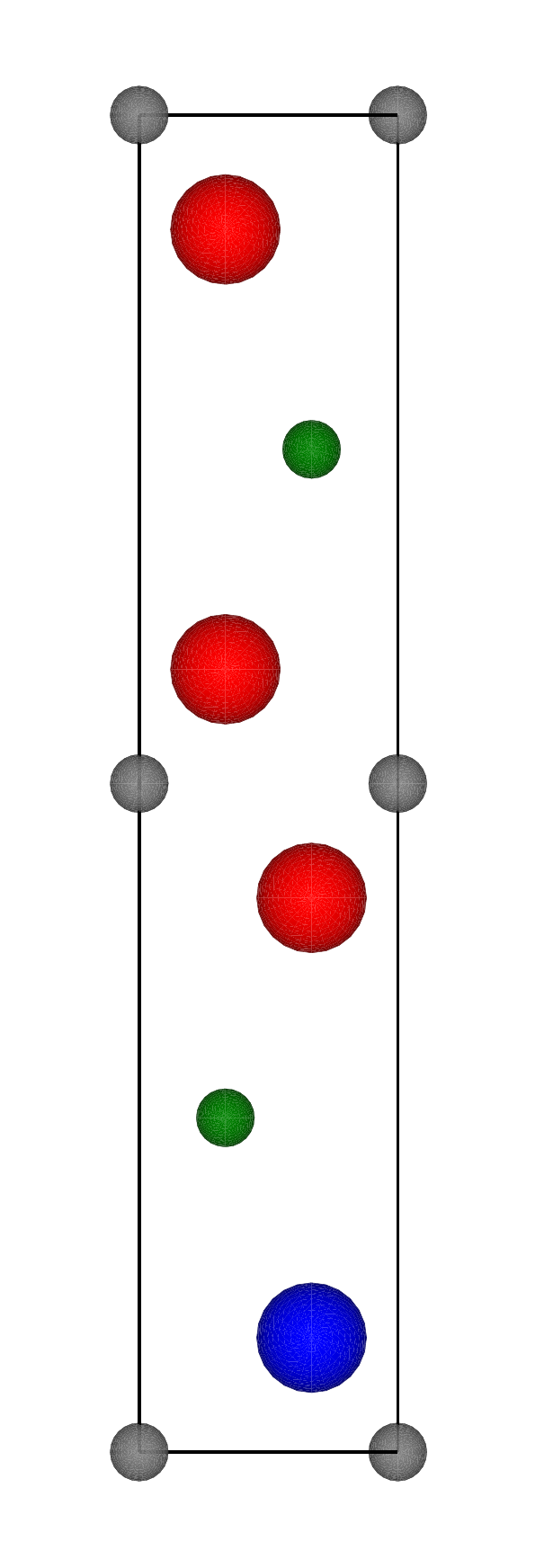}
  VVVCr
  \vspace{3mm}
\endminipage
&
\minipage{0.11\textwidth}
  \centering
  \includegraphics[width=\linewidth]{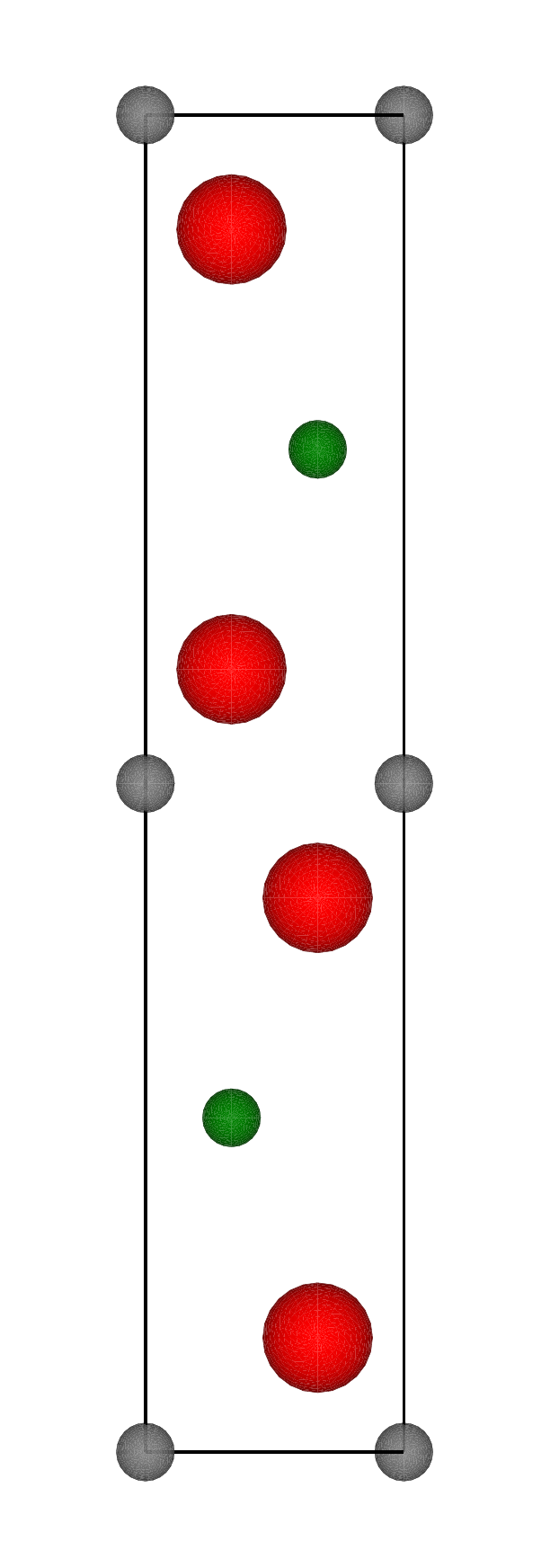}
  V${_{2}}$AlC
  \vspace{3mm}
\endminipage \\ 

& 

& 
\minipage{0.025\textwidth}
  \centering	
  \includegraphics[width=\linewidth]{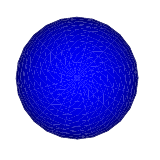}
\endminipage
\,\,\,\,
\minipage{0.025\textwidth}
  \centering	
  \includegraphics[width=\linewidth]{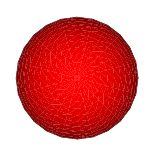}
\endminipage 
\,\,\,\,
\minipage{0.0145\textwidth}
  \centering	
  \includegraphics[width=\linewidth]{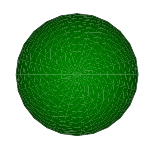}    
\endminipage
\,\,\,\,
\minipage{0.0135\textwidth}
  \centering	
  \includegraphics[width=\linewidth]{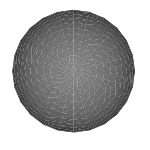} 
\endminipage
& 

& \\ 

& & Cr\,\,\,\,\,\,\,\,\,V\,\,\,\,\,\,\,Al\,\,\,\,\,C & & \\  
  \end{tabular}
\caption{Single 1x1x1 unit cells of the ordered structures considered for $\text{(}\text{Cr}_\text{x}\text{,}\text{V}_{\text{1-x}}\text{)}_{\text{2}}\text{AlC}$. The vertical direction points along the c-axis.}\label{tbl:schematic-unit-cell} 
\end{table*}

The present contribution deals with chemically ordered and disordered quaternary $\text{(}\text{Cr}_\text{x}\text{,}\text{V}_{\text{1-x}}\text{)}_{\text{2}}\text{AlC}$ phases for ${x}$ = 0; 0.25; 0.5; 0.75; 1.0. These phases are normally synthesized at temperatures between 1673 K and 1873 K\cite{Halim2017} by reactive sintering or hot isostatic pressing. This work aims to provide a theoretical evaluation on the phase stability and magnetic properties of these alloys using first principles calculations.

\section{Method}

${Ab\,\,\,initio\,}$ calculations based on density functional theory\cite{Hohenberg1964,Kohn1965} were performed using the SIESTA program\cite{Soler2002}. We adopted the generalized gradient approximation (GGA) as parameterised by Perdew-Burke-Ernzerhof (PBE)\cite{Perdew1996} for treating electron exchange and correlation effects. All structural relaxations were done using the conjugate gradient method\cite{Hestenes1952}, to within a force tolerance of $\text{0.02 eV/\AA}$ and a stress tolerance of 0.01 GPa. Integrations in real space were performed using a real-space grid with a 300 Ry mesh cutoff. For the k-point sampling of the Brillouin zone, values between $\text{20-25\AA}$ were used for the k-grid cutoff length\cite{Moreno1992} depending on each alloy.

Core electrons were replaced by pseudopotentials\cite{Soler2002}. Two different valence configurations were considered in order to generate Cr and V pseudopotentials: the first configuration (${fs}$) taken as Cr(${\text{3}\text{p}^{\text{6}}\text{3}\text{d}^{\text{5}}\text{4}\text{s}^{\text{1}}}$), V(${\text{3}\text{p}^{\text{6}}\text{3}\text{d}^{\text{3}}\text{4}\text{s}^{\text{2}}}$) and the second configuration (${ss}$) taken as Cr(${\text{3}\text{s}^{\text{2}}\text{3}\text{p}^{\text{6}}\text{3}\text{d}^{\text{5}}}$), V(${\text{3}\text{s}^{\text{2}}\text{3}\text{p}^{\text{6}}\text{3}\text{d}^{\text{3}}}$). For Al and C pseudopotential generation, valences were taken as Al(${\text{3}\text{s}^{\text{2}}\text{3}\text{p}^{\text{0.5}}\text{3}\text{d}^{\text{0.5}}}$), C(${\text{2}\text{s}^{\text{2}}\text{2}\text{p}^{\text{2}}}$). In addition, Kohn-Sham eigenvectors were expanded in atomic-like orbitals and basis sets were chosen to be the solutions of the pseudo-atomic problem. Three basis sets were defined for Cr and V according to the valence configurations given above and denoted DZP, DZP+P(3d) and TZ2P+P(3d). For Al and C, a DZP basis set was chosen for each. More detailed information on the basis sets and pseudopotentials can be found in Appendix A.

Figure 1 shows single unit cells of $\text{(}\text{Cr}_\text{x}\text{,}\text{V}_{\text{1-x}}\text{)}_{\text{2}}\text{AlC}$ (${x}$ = 0; 0.25; 0.5; 0.75; 1.0) for structures which are chemically ordered in layers along the c direction. Each unit cell contains 8 atoms. To simulate the effects of M-site disorder, two further cells were considered focusing on the equiatomic composition CrVAlC (${x}$ = 0.5).  One was modelled using the special quasirandom structure method\cite{Zunger1990} using a 4x4x1 supercell containing 64 M-sites and denoted SQSCrV. The other was modelled using a 2x1x1 supercell with a total of 8 M-sites and denoted VVCrCr\textbf{(}CrV\textbf{)}\textbf{(}VCr\textbf{)}. This structure is shown in Table B1 of Appendix B.

Non-magnetic (NM) calculations (following the nomenclature of previous papers on
the topic) refer to spin-paired or non-spin-polarised DFT calculations.
They should not be confused with paramagnetism, which allows for the presence of 
disordered magnetic moments. 
DFT calculations allowing for magnetic moments (spin polarised), are performed in this work 
for various spin arrangements, for ${x}$ = 0, 0.5 and 1.
  The different arrangements found as most stable for the different mixing configurations
explored in this work all give zero total magnetisation, and will be generally referred to as 
AFM following the literature, although the structure defined by the Cr atoms is never 
bipartite and all configurations contain a certain degree of  frustration.
  The particular arrangements obtained are not trivial, and are specified for ${x}$ = 0.5 in Table B1 
of Appendix B.
  The most stable spin configuration for the Cr end member coincides with what found in 
previous works on that material\cite{Dahlqvist2013,2Dahlqvist2015}, where it is called
in-AFM1.
  Ferromagnetic (FM) configurations were also obtained but always found to be of significantly
higher energy, and are therefore not reported in this work.  
  Finally, the rotationally invariant approach to GGA${+U}$ as proposed by Dudarev\cite{Dudarev1998} was applied to describe the Cr 3$d$ electrons 
(see Appendix B for the parameters used in the definition of the Hamiltonian).

The Gibbs free energy of mixing ${\Delta G_{\text{mix}}}$ for each alloy is expressed as
\begin{equation}
\Delta G_{\text{mix}}= \Delta H_{\text{mix}} -T\Delta S_{\text{mix}}
\end{equation}
where ${\Delta H_{\text{mix}}}$ is the enthalpy of mixing given by
\begin{equation}
\Delta H_{\text{mix}}= \Delta E_{\text{mix}} + p\Delta V_{\text{mix}}  
\end{equation}
for a given pressure $p$, and where $\Delta S_{\text{mix}}$, $\Delta E_{\text{mix}}$, 
and $\Delta V_{\text{mix}}$ 
are the mixing entropy, internal energy, and volume, respectively, each one of them defined as
\begin{equation}
\Delta \Omega_{\text{mix}} =  \Omega_{\text{mixed}} - \Omega_{\text{pure1}} 
-\Omega_{\text{pure2}}
\end{equation}
for $\Omega = S$, $E$, and $V$, respectively, all of them expressed per unit cell 
henceforth.

We expect the configurational entropy ${\Delta S_{\text{c}}}$ to be the main component of the total entropy of mixing, since changes in vibrational entropy are expected to be small in comparison: the replacement of one M-site transition-metal atom by another with very similar mass and chemistry
should not change the phonon frequencies of the different phases enough so as to significantly affect
the mixing entropy. We also expect configurational entropy to dominate over spin disorder entropy
in $\Delta S_{\text{mix}}$, since, for the former, the end members give exactly zero contribution, 
maximising the mixing total, which is not the case for the latter. 

The internal energy of mixing will be calculated at 0 K in this work since the variation with temperature is also expected to be sufficiently small to be negligible for present purposes. Finally, concerning the $pV$ term, there are two synthesis processes for these materials\cite{Halim2017}, one is at 0.1 MPa and the other one involves a pressure of 80 MPa which makes ${p\Delta V}$ less than 2 meV per unit cell for a ${\Delta V \textless \,\, 3 \,\, \%}$ in all cases.
Thus the resulting expression used to calculate the enthalpy of mixing for each alloy was
\begin{eqnarray}
\Delta H_{\text{mix}}&=& E(\text{(}\text{Cr}_\text{x}\text{,}\text{V}_{\text{1-x}}\text{)}_{\text{2}}\text{AlC}) -x\,E(\text{Cr}_{\text{2}}\text{AlC}) \nonumber\\& &-(1-x)\,E(\text{V}_{\text{2}}\text{AlC})  
\end{eqnarray}
where ${E}$ is the internal energy of the corresponding phase.\\

The configurational entropy ${\Delta S_{\text{c}}}$ per unit cell of an ideal solid solution of V and Cr atoms on the M-sites has the form
\begin{equation}
\Delta S_{\text{c}}=-y \, k_{\text{B}}\, \text{[}{x}\,\, \text{ln}\text{(}{x}\text{)}+ \text{(}\text{1-}x\text{)} \,\,\text{ln}\text{(}\text{1-}x\text{)}\text{]} ,
\end{equation}
where ${y}$ is the number of M-sites per unit cell, i.e., ${y}$ = 4 and ${x}$ is the Cr molar content. When ${x}$ = 0.5, for example, ${\Delta S_{\text{c}}=-23.89 x 10^{-5}}$ eV/K. Deviations from
the ideal mixing behaviour are not considered in this work.

The following section describes results obtained using the pseudopotentials associated to the ${fs}$ valence configurations and DZP+P(3d) basis sets for Cr and V. Appendix A shows that using the ${ss}$ valence configurations or the other basis sets has only a small effect on the enthalpy of mixing.

\setcounter{figure}{1}
\begin{figure}
\begin{center}
  \includegraphics[width=0.5\textwidth]{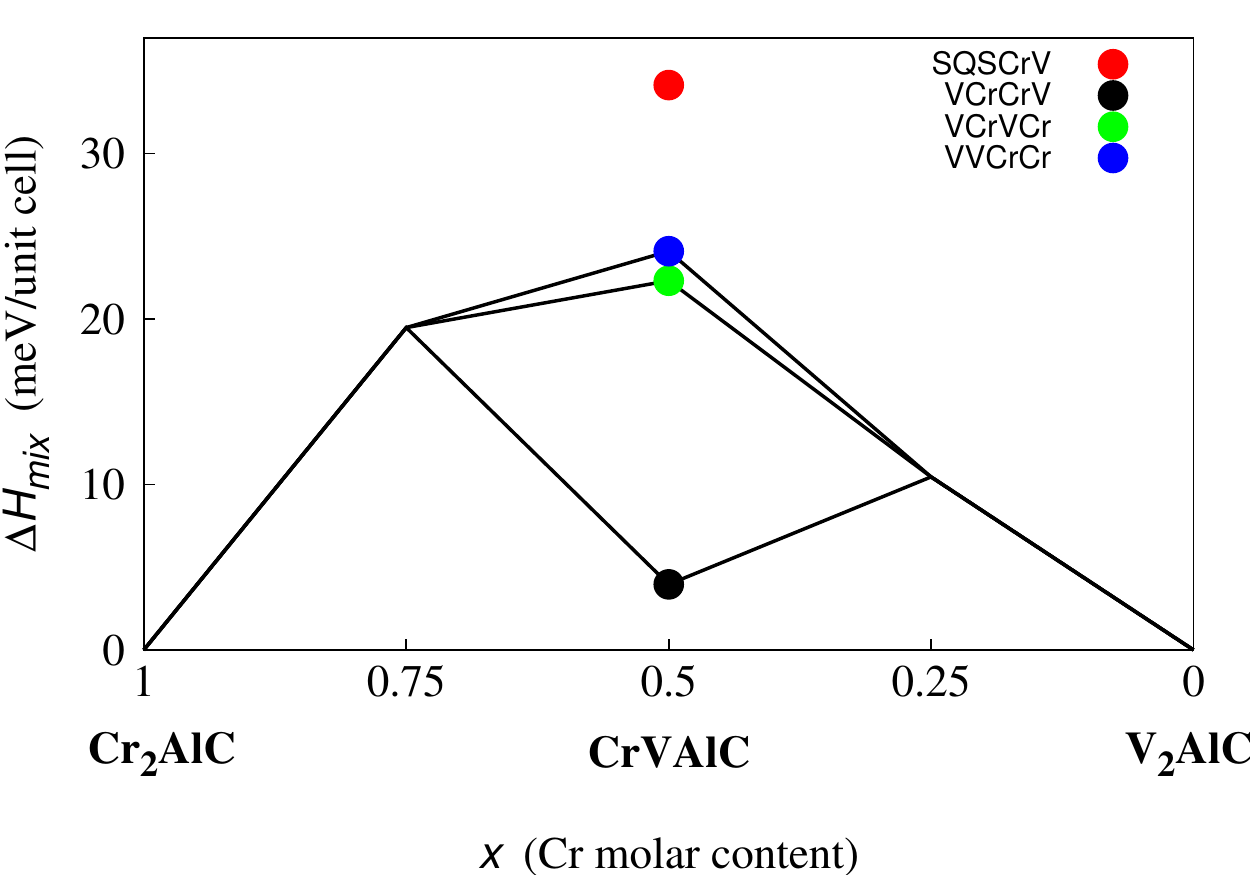}
\end{center}
\caption{Enthalpy of mixing ${\Delta H_{\text{mix}}}$ for the chemically ordered structures of $\text{(}\text{Cr}_\text{x}\text{,}\text{V}_{\text{1-x}}\text{)}_{\text{2}}\text{AlC}$ shown in FIG.1 as a function of ${\,\,x}$. The enthalpy for the disordered structure SQSCrV at ${x}$ = 0.5 is also shown. All configurations have been taken to be NM.}\label{fig:physics} 
\end{figure}

\section{Results}

Figure 2 shows ${\Delta H_{\text{mix}}}$ as a function of composition for the chemically ordered structures with ${x}$ = 0; 0.25; 0.5; 0.75; 1.0. Also plotted, for comparison, is ${\Delta H_{\text{mix}}}$ for the disordered SQSCrV structure at  ${x}$ = 0.5.  All configurations have been taken to be nonmagnetic. It is seen that the enthalpies are all positive resulting in a concave hull and therefore indicating that each configuration is unstable with respect to decomposition into the end members at 0 K. At ${x}$ = 0.5, VCrCrV is the least unstable structure whereas SQSCrV is the most unstable structure.

To determine the effect of magnetism and magnetic order on ${\Delta H_{\text{mix}}}$ we have performed spin polarised calculations focusing on the ${x}$ = 0.5 equiatomic composition. In addition, DFT+${U}$ has been employed to see whether the on-site Coulomb interactions of the localised 3d electrons of Cr influence the results. In previous studies\cite{Dahlqvist2013,2Dahlqvist2015} an antiferromagnetic state (in-AFM1) was found to be the ground state for $\text{Cr}_{\text{2}}\text{AlC}$ and the present calculations confirm this result. The ground state for $\text{V}_{\text{2}}\text{AlC}$ was found to be NM\cite{Schneider2006}. We tested this by starting with the same spin options proposed for $\text{Cr}_{\text{2}}\text{AlC}$ and found that  $\text{V}_{\text{2}}\text{AlC}$ relaxes into the NM state in all cases. For the equiatomic composition, three different magnetic orders (AFM, FM and NM) have been considered. The AFM state is always preferred over the FM state.  Details of the preferred AFM spin configurations for the chemically ordered (VCrCrV, VCrVCr and VVCrCr) and disordered (VVCrCr(CrV)(VCr) structures are given in Appendix B. 
Figure 3 compares the formation enthalpies ${\Delta H_{\text{mix}}}$ of the AFM configurations with the NM configurations at ${x}$ = 0.5. For the AFM calculations, both DFT and DFT+${U}$ were considered, the latter using a value of ${U}$ = 1 eV following previous work on Cr\cite{2Dahlqvist2015}. The NM results reproduce the values given in FIG. 2. It is clearly seen that the alloys become more unstable when magnetism is considered. Although magnetism and ${+U}$ corrections tend to favour different chemically ordered structures, the disordered arrangements (SQSCrV for NM and VVCrCr(CrV)(VCr) for AFM and AFM+ ${U}$) always have the highest enthalpies and are the most unstable.

\setcounter{figure}{2}
\begin{figure}[!htb]
\begin{center}
\minipage{0.5\textwidth}
  \includegraphics[width=\linewidth]{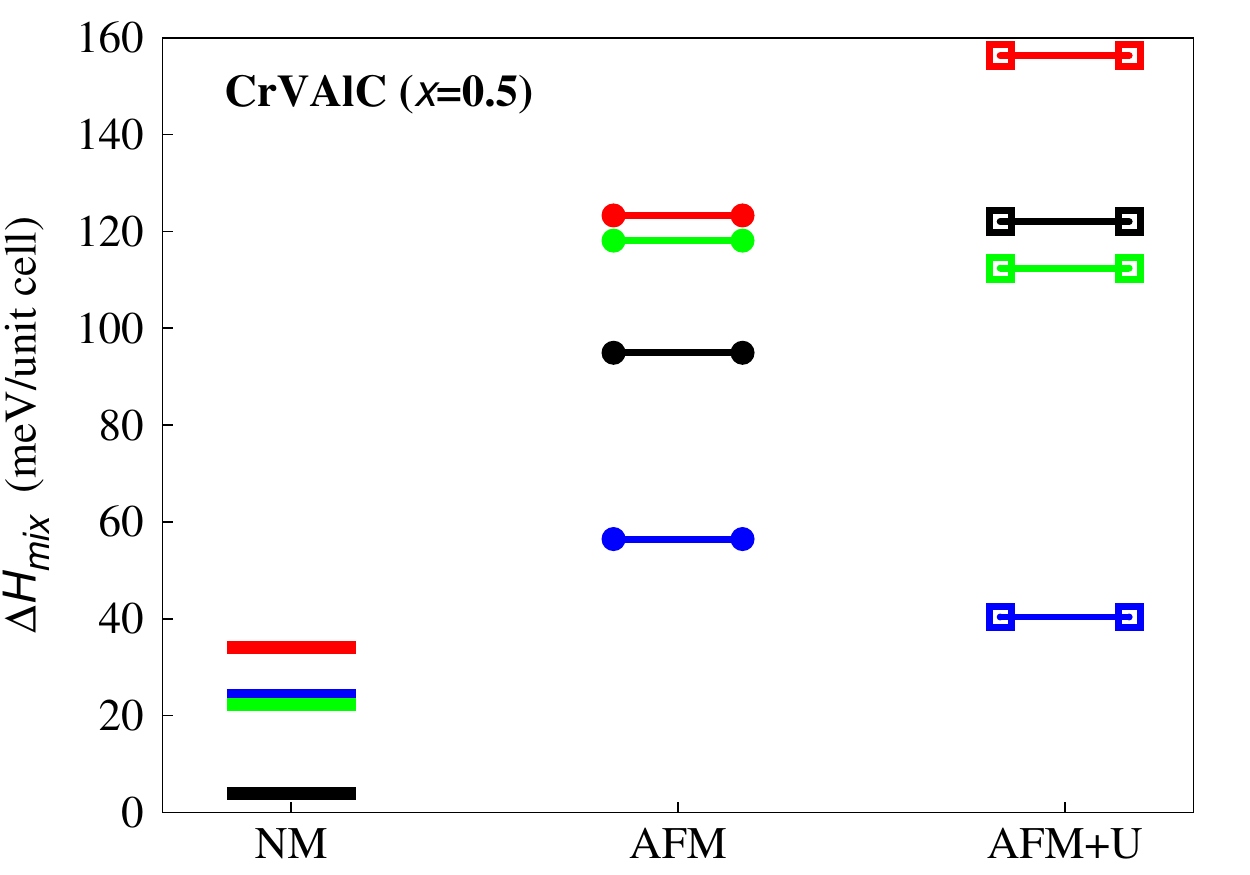}
\endminipage
\end{center}
\caption{Enthalpy of mixing ${\Delta H_{\text{mix}}}$ for the NM, AFM and AFM${+U}$ (${U = \text{1 eV }}$) configurations of CrVAlC (${x}$ = 0.5). Colour code: SQSCrV (NM, red), VVCrCr\textbf{(}CrV\textbf{)}\textbf{(}VCr\textbf{)} (AFM and AFM+${U}$, red), VCrCrV (black), VCrVCr (green) and VVCrCr (blue). Lines without symbols: NM; with circles: AFM; with squares: AFM${+U}$.}\label{fig:fmu_afmu}
\end{figure}

\renewcommand{\tablename}{FIG.}
\setcounter{table}{3}
\begin{table}[!h]
\begin{tabular}{c}
\begin{minipage}{0.487\textwidth}
  \includegraphics[width=\linewidth]{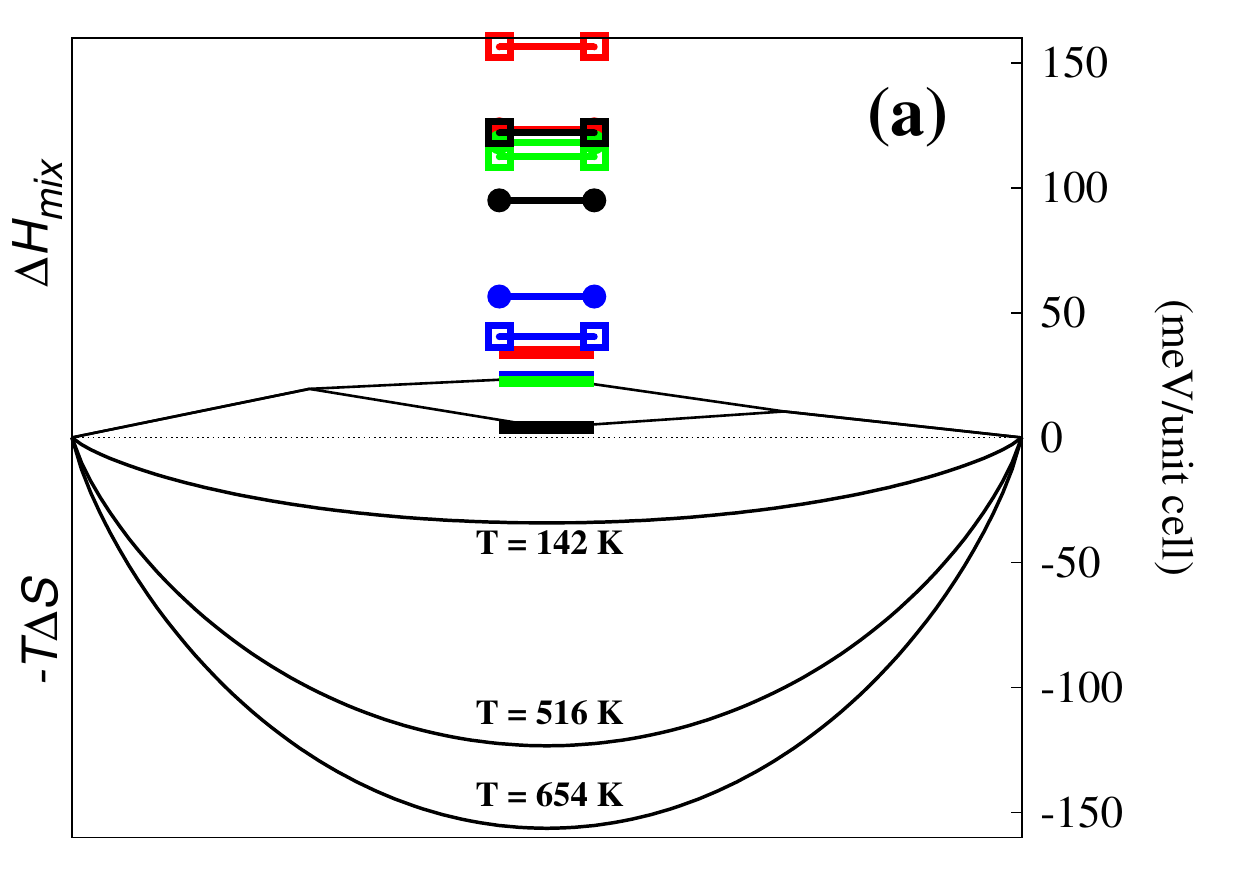}
\end{minipage}
\\
\begin{minipage}{0.49\textwidth}
  \vspace{-0.4cm}
  \includegraphics[width=\linewidth]{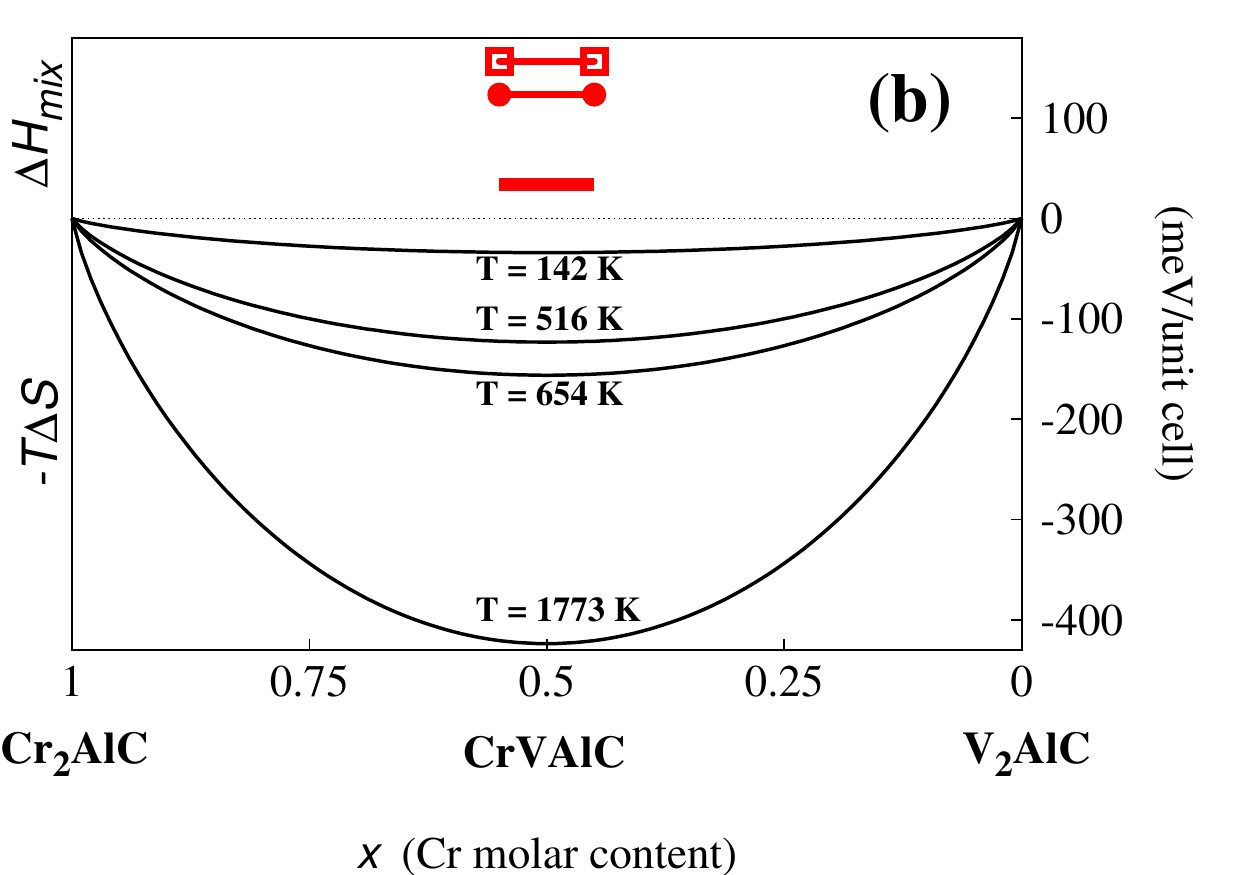}
\end{minipage}
\end{tabular}  
\caption{\textbf{(a)} Enthalpy of mixing ${\Delta H_{\text{mix}}}$ for all chemically ordered and disordered structures of $\text{(}\text{Cr}_\text{x}\text{,}\text{V}_{\text{1-x}}\text{)}_{\text{2}}\text{AlC}$ considered in this work as a function of ${\,\,x}$. The same colour coding is used as in Figs. 2 and 3. Also shown are the ${-T\Delta S_{\text{c}}}$ curves for three equiatomic (${x}$ = 0.5) structures at the critical temperature when ${T\Delta S_{\text{c}}}$ is equal to ${\Delta H_{\text{mix}}}$. The chosen structures, magnetic configurations and temperatures are SQSCrV (NM, T = 142 K), VVCrCr\textbf{(}CrV\textbf{)}\textbf{(}VCr\textbf{)} (AFM, T = 516 K) and VVCrCr\textbf{(}CrV\textbf{)}\textbf{(}VCr\textbf{)} (AFM${+U}$, T = 654 K).
\textbf{(b)} Same as {(a)} except only disordered structures (with the associated ${-T\Delta S_{\text{c}}}$ curves) are shown along with the curve for the mean experimental temperature (T=1773 K).} \label{fig:entropy1-2} 
\end{table}

Figure 4(a) re-plots the enthalpies of mixing shown in Figs. 2 and 3 but also includes the ${-T\Delta S_{\text{c}}}$ curves for three equiatomic (${x}$ = 0.5) structures for the specific values of the temperature such that ${T\Delta S_{\text{c}}}={\Delta H_{\text{mix}}}$ for the NM, AFM and
AFM$+U$ results, giving estimates of the stabilisation temperature $T_s$ for which 
${\Delta G_{\text{mix}} = 0}$, and below which the mixed phase is therefore thermodynamically 
unstable. 
The chosen structures and magnetic configurations for the different cases are: SQSCrV (NM), VVCrCr\textbf{(}CrV\textbf{)}\textbf{(}VCr\textbf{)} (AFM) and VVCrCr\textbf{(}CrV\textbf{)}\textbf{(}VCr\textbf{)} (AFM${+U}$) because these are the most unstable phases for this composition in each case (red lines in Fig. 3).  The corresponding stabilisation temperatures are $T_s = 142$~K, 516~K and 654~K, respectively. 
Figure 4(b) focuses only on the $x = 0.5$ composition and includes the  ${-T\Delta S_{\text{c}}}$ curve corresponding to the mean experimental temperature $T = 1773$~K\cite{Halim2017}.   
Figure 5 shows ${\Delta G_{\text{mix}}}$ plotted as a function of temperature for the three equiatomic structures. At T = 0 K the expression ${\Delta G_{\text{mix}} = \Delta H_{\text{mix}}}$ is recovered for the corresponding values of the enthalpy of mixing. When ${\Delta G_{\text{mix}} = 0}$ the 
expression ${\Delta H_{\text{mix}} = T\Delta S_{\text{mix}}}$ is obtained together with the critical temperatures.

\setcounter{figure}{4}
\begin{figure}[!htb]
\begin{center}
\minipage{0.5\textwidth}
  \includegraphics[width=\linewidth]{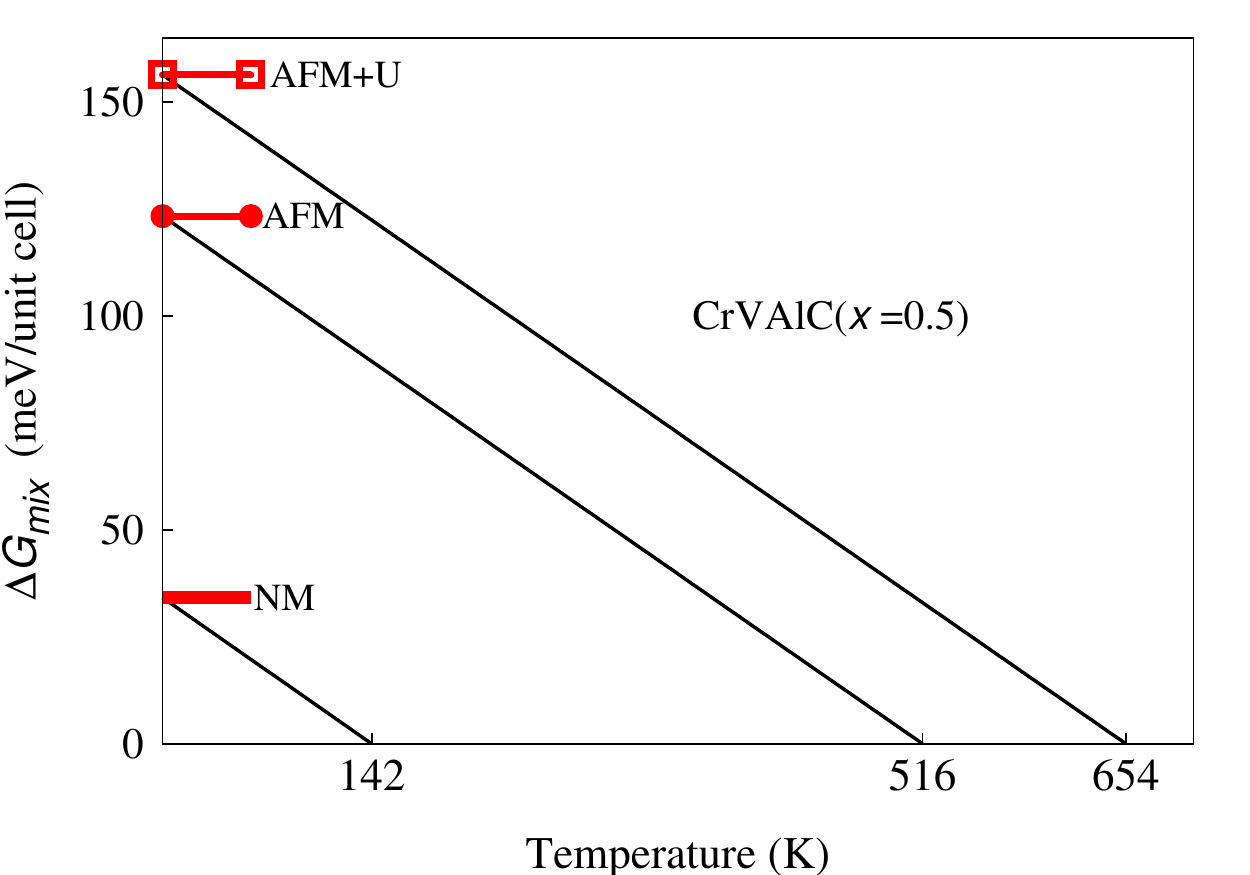}
\endminipage
\end{center}
\caption{Gibbs free energy of mixing ${\Delta G_{\text{mix}}}$ for NM, AFM and AFM${+U}$ configurations of CrVAlC (${x}$ = 0.5) as a function of temperature. ${\Delta G_{\text{mix}}}$  becomes zero at the corresponding critical temperatures.}\label{fig:entropy3}
\end{figure}

It is thus predicted that the mixed phase $\text{(}\text{Cr}_\text{0.5}\text{,}\text{V}_{\text{0.5}}\text{)}_{\text{2}}\text{AlC}$ is thermodynamically stable at temperatures higher than a stabilisation 
temperature $T_s$ estimated to be around 600 K.
  The DFT$+U$ estimate is considered here to be probably the most accurate
obtained in this work, since it does show well defined magnetic moments for 
the Cr atoms, but higher levels of theory (beyond DFT, when amenable) should give more 
accurate estimates.  
It should be remembered however, that a more precise definition of $T_s$ would require the 
explicit consideration of finite-temperature corrections to $\Delta E_{\text{mix}}$, the inclusion 
of $p\Delta V_{\text{mix}}$ terms, and a more accurate calculation of $\Delta S_{\text{mix}}$
including vibrational contributions as well as correlation corrections to the configurational entropy, 
and spin-disorder. It is clear, however, that, in spite of the low magnetic transition temperature
measured for the Cr end member\cite{Jaouen2013} of 73 K, which would seem to imply that
magnetism would not be relevant at significantly higher temperatures, a NM solution would give 
a rather poor estimate of $T_s=142$ K, although reaching the same qualitative conclusion. 

It is therefore expected that the mixed phases are stable at the synthesis conditions (as 
observed), and remain metastable when cooled down to room temperature for kinetic reasons. 
This scenario would also predict that if $T_s$ is sufficiently high to allow significant diffusion
when cooling well below this temperature, a tendency towards spinodal decomposition
would exist, which might be observed if cooling sufficiently slowly. 
An estimation of the time scales relevant for this phenomenon to be observed would 
require a more precise knowledge of $T_s$, and good estimates of the relevant
diffusivities, which are beyond the scope of this work. For the configurational disorder that remains frozen-in, however, the antiferromagnetic frustration already seen in the Cr end member (due to the hexagonal Cr layers) could give rise to interesting spin orderings when the Cr cations alternate with non-magnetic V cations in a disordered fashion. The spin configurations shown in Table B1 are just a 
small sample of what less ordered cation arrangements could produce, including ferrimagnetic response, plausibly involving spin canting.

\section{Conclusions}

Using first principles DFT calculations, the effect of temperature and magnetism on the stability of $\text{(}\text{Cr}_\text{x}\text{,}\text{V}_{\text{1-x}}\text{)}_{\text{2}}\text{AlC}$ MAX phases has been studied. At T = 0 K, calculations of the enthalpy of mixing indicate that chemically ordered structures across the composition range are unstable with respect to decomposition into the two end members. Further calculations at the equiatomic composition, CrVAlC (${x}$ = 0.5), show that the effect of chemical disorder does not change this conclusion, but tend to make the structures even more unstable. Calculations including different magnetic arrangements uphold this conclusion. However, when configurational entropy is included, these disordered structures can be stabilised with temperature. It is found that at the equiatomic composition, a NM configuration of the disordered structure SQSCrV becomes stable at $T_s = 142$ K. Introducing magnetism in AFM configurations  and changing the nature of the chemical disorder (VVCrCr\textbf{(}CrV\textbf{)}\textbf{(}VCr\textbf{)}) increases the stabilisation temperature to $T_s = 516$ K - 654 K depending on the level of theory used (DFT or DFT${+U}$). Thus, although the measured Curie temperature of $\text{Cr}_{\text{2}}\text{AlC}$ is about 73 K, the calculations predict that introducing vanadium can not only stabilise the quaternary phase at temperatures well below those typically used during synthesis, but also induce new magnetic arrangements. It is hoped that the present work will stimulate new measurements of the magnetic properties of the $\text{(}\text{Cr}_\text{x}\text{,}\text{V}_{\text{1-x}}\text{)}_{\text{2}}\text{AlC}$ solid solution series.

\begin{acknowledgments}
The authors gratefully acknowledge the computational resources from Marenostrum III and IV (Barcelona Supercomputer Center, Spain). EA thanks EC-FP7-PEOPLE-CIG-2012 Marie Curie, Project number 333813 ElectronStopping of the European Union. PDB and SHS thank the Donostia International Physics Center (DIPC) and EPSRC Grant EP/M018768/1 respectively for travel and subsistence during their visits.  
\end{acknowledgments}

\appendix

\section*{Appendix A}

Pseudopotentials for Cr and V have been generated using the program ATOM\cite{Garcia2006} considering the relativistic PBE\cite{Perdew1996} functional within the Troullier-Martins (tm2)\cite{Troullier1991} scheme. Basis sets for Cr and V, denoted DZP, DZP+P(3d) and TZ2P+P(3d), have been defined according to the valence configurations ${fs}$ and ${ss}$. The parameters for Cr, V, Al and C used in this work to generate the corresponding pseudopotentials are given in Table A1 while those for the basis sets are given in Tables A2 and A3.

\renewcommand{\thetable}{A1}
\renewcommand{\tablename}{TABLE}
\setcounter{table}{0}
\begin{table}[!h]
\caption{Parameters for generating the Cr and V pseudopotentials for the two valence configurations ${fs}$ and ${ss}$. The parameters used to generate the Al and C pseudopotentials are also given. ${\text{r}_{\text{c}}\text{(s, p, d or f)}}$ refer to the core radii for the orbitals (values are in Bohr) and s, d, p and f are the corresponding electron occupancies. ${\text{r}_{\text{core}}}$ is the radius for including partial core corrections\cite{Louie1982}.}\label{tbl:data for pseudo generation}
\begin{ruledtabular}
  \begin{tabular}{c|c|ccccc|cccc}
            & & ${\text{r}_{\text{c}}\text{(s)}}$ & ${\text{r}_{\text{c}}\text{(p)}}$ & ${\text{r}_{\text{c}}\text{(d)}}$ & ${\text{r}_{\text{c}}\text{(f)}}$ & ${\text{r}_{\text{core}}}$ & s & p & d & f \\ \hline
    ${fs}$ & Cr & 2.80 & 1.70 & 2.50 & 2.25 & 1.85 & 1.0 & 6.0 & 5.0 & 0.0 \\
            & V & 2.80 & 1.65 & 2.50 & 2.25 & 1.74 & 2.0 & 6.0 & 3.0 & 0.0 \\ \hline
    ${ss}$ & Cr & 1.00 & 1.60 & 2.50 & 2.25 & 1.38 & 2.0 & 6.0 & 5.0 & 0.0 \\
            & V & 1.00 & 1.60 & 2.50 & 2.25 & 1.38 & 2.0 & 6.0 & 3.0 & 0.0 \\ \hline
            & Al & 2.28 & 2.28 & 2.28 & 2.28 & 1.5 & 2.0 & 0.5 & 0.5 & 0.0 \\
            & C & 1.25 & 1.25 & 1.25 & 1.25 & 1.6 & 2.0 & 2.0 & 0.0 & 0.0 \\           
  \end{tabular}
\end{ruledtabular}
\end{table}
Figure A1 compares the calculated enthalpy of mixing of the NM state for the VVCrCr structure using the different basis sets and pseudopotentials. The physical situation remains the same regardless of the particular choice of basis or pseudopotential. The largest value of ${\Delta H_{\text{mix}}}$ over the composition range is obtained using the ${ss}$ valence configuration and the DZP+P(3d) basis set. There is a difference of approximately 5 meV/unit cell between this choice and the one using the ${fs}$ configuration (same basis). This difference is not large enough to have a significant effect on the overall results.

\renewcommand{\thetable}{A2}
\renewcommand{\tablename}{TABLE}
\setcounter{table}{0}
\begin{table*}[t]
 \caption{Parameters for the basis sets DZP, DZP+P(3d) and TZ2P+P(3d) associated with the two valence configurations defined for Cr and V. ${\text{r}_{\text{1, 2 or 3}}}$ are cutoff radii (Bohr) of each ``zeta'' for the orbital (orb). ${\text{r}_{\ast}}$ is the internal radius of the soft confinement potential\cite{Junquera2001}(with ${\text{V}_{0} \text{= 40 Ry}}$) while ${\text{r}_{\text{Q}}}$ is the radius associated to the charge confinement potential\cite{Corsetti2013}.}\label{tbl:Cr-V basis}
\begin{ruledtabular}
 \begin{tabular}{c|cccccc|cccccc|cccccc|cccccc}
   & \multicolumn{12}{c|}{Cr Basis} & \multicolumn{12}{c}{V Basis} \\ \hline
   & & & ${fs}$ & & & & & & ${ss}$ & & & & & & ${fs}$ & & & & & & ${ss}$ & & &\\ \hline
   & orb & ${\text{r}_{\text{1}}}$ & ${\text{r}_{\text{2}}}$ & ${\text{r}_{\text{3}}}$ & ${\text{r}_{\ast}}$ & ${\text{r}_{\text{Q}}}$ & orb & ${\text{r}_{\text{1}}}$ & ${\text{r}_{\text{2}}}$ & ${\text{r}_{\text{3}}}$ & ${\text{r}_{\ast}}$ & ${\text{r}_{\text{Q}}}$ & orb & ${\text{r}_{\text{1}}}$ & ${\text{r}_{\text{2}}}$ & ${\text{r}_{\text{3}}}$ & ${\text{r}_{\ast}}$ & ${\text{r}_{\text{Q}}}$ & orb & ${\text{r}_{\text{1}}}$ & ${\text{r}_{\text{2}}}$ & ${\text{r}_{\text{3}}}$ & ${\text{r}_{\ast}}$ & ${\text{r}_{\text{Q}}}$\\ \hline 
DZP & ${3_{\text{p}}}$ & 10.0 & - & - & 8.5 & - & ${3_{\text{s}}}$ & 10.0 & 2.5 & - & 8.5 & - & ${3_{\text{p}}}$ & 10.0 & - & - & 8.5 & - & ${3_{\text{s}}}$ & 10.0 & 3.0 & - & 8.5 & -  \\
    & ${3_{\text{d}}}$ & 10.0 & 3.0 & - & 8.5 & - & ${3_{\text{p}}}$ & 10.0 & 2.5 & - & 8.5 & - & ${3_{\text{d}}}$ & 10.0 & 3.0 & - & 8.5 & - & ${3_{\text{p}}}$ & 10.0 & 3.0 & - & 8.5 & - \\ 
    & ${4_{\text{s}}}$ & 10.0 & 3.0 & - & 8.5 & - & ${3_{\text{d}}}$ & 10.0 & 3.0 & - & 8.5 & - & ${4_{\text{s}}}$ & 10.0 & 3.0 & - & 8.5 & - & ${3_{\text{d}}}$ & 10.0 & 3.5 & - & 8.5 & - \\
    & ${4_{\text{p}}}$ & 10.0 & - & - & 8.5 & 1.3 & ${4_{\text{s}}}$ & 10.0 & 3.0 & - & 8.5 & - & ${4_{\text{p}}}$ & 10.0 & - & - & 8.5 & 1.3 & ${4_{\text{s}}}$ & 10.0 & 3.5 & - & 8.5 & - \\
    & & & & & & & ${4_{\text{p}}}$ & 10.0 & - & - & 8.5 & 1.38 & & & & & & & ${4_{\text{p}}}$ & 10.0 & - & - & 8.5 & 1.38 \\ \hline   
DZP+P(3d) & ${3_{\text{p}}}$ & 10.0 & - & - & 8.5 & - & ${3_{\text{s}}}$ & 10.0 & 2.5 & - & 8.5 & - & ${3_{\text{p}}}$ & 10.0 & - & - & 8.5 & - & ${3_{\text{s}}}$ & 10.0 & 3.0 & - & 8.5 & - \\
    & ${3_{\text{d}}}$ & 10.0 & 3.0 & - & 8.5 & - & ${3_{\text{p}}}$ & 10.0 & 2.5 & - & 8.5 & - & ${3_{\text{d}}}$ & 10.0 & 3.0 & - & 8.5 & - & ${3_{\text{p}}}$ & 10.0 & 3.0& - & 8.5 & - \\ 
    & ${4_{\text{f}}}$ & 10.0 & - & - & - & - & ${3_{\text{d}}}$ & 10.0 & 3.0 & - & 8.5 & - & ${4_{\text{f}}}$ & 10.0 & - & - & - & - & ${3_{\text{d}}}$ & 10.0 & 3.5 & - & 8.5 & - \\ 
    & ${4_{\text{s}}}$ & 10.0 & 3.0 & - & 8.5 & - & ${4_{\text{f}}}$ & 10.0 & - & - & - & - & ${4_{\text{s}}}$ & 10.0 & 3.0 & - & 8.5 & - & ${4_{\text{f}}}$ & 10.0 & - & - & - & - \\
    & ${4_{\text{p}}}$ & 10.0 & - & - & 8.5 & 1.3 & ${4_{\text{s}}}$ & 10.0 & 3.0 & - & 8.5 & - & ${4_{\text{p}}}$ & 10.0 & - & - & 8.5 & 1.3 & ${4_{\text{s}}}$ & 10.0 & 3.5 & - & 8.5 & - \\
    & & & & & & & ${4_{\text{p}}}$ & 10.0 & - & - & 8.5 & 1.38 & & & & & & & ${4_{\text{p}}}$ & 10.0 & - & - & 8.5 & 1.38 \\ \hline        
TZ2P+P(3d) & ${3_{\text{p}}}$ & 10.0 & 4.0 & 3.0 & 9.0 & - & ${3_{\text{s}}}$ & 10.0 & 3.5 & 2.5 & 9.0 & - & ${3_{\text{p}}}$ & 10.0 & 4.0 & 3.0 & 9.0 & - & ${3_{\text{s}}}$ & 10.0 & 4.0 & 2.5 & 9.0 & - \\
    & ${3_{\text{d}}}$ & 10.0 & 4.0 & 3.0 & 9.0 & - & ${3_{\text{p}}}$ & 10.0 & 3.5 & 2.5 & 9.0 & -  & ${3_{\text{d}}}$ & 10.0 & 4.0 & 3.0 & 9.0 & - & ${3_{\text{p}}}$ & 10.0 & 4.0 & 2.5 & 9.0 & - \\ 
    & ${4_{\text{f}}}$ & 10.0 & - & - & - & - & ${3_{\text{d}}}$ & 10.0 & 4.0 & 3.0 & 9.0 & - & ${4_{\text{f}}}$ & 10.0 & - & - & - & - & ${3_{\text{d}}}$ & 10.0 & 4.5 & 3.5 & 9.0 & - \\ 
    & ${4_{\text{s}}}$ & 10.0 & 4.0 & 3.0 & 9.0 & - & ${4_{\text{f}}}$ & 10.0 & - & - & - & - & ${4_{\text{s}}}$ & 10.0 & 4.0 & 3.0 & 9.0 & - & ${4_{\text{f}}}$ & 10.0 & - & - & - & - \\
    & ${4_{\text{p}}}$ & 10.0 & 3.0 & - & 9.0 & 1.3 & ${4_{\text{s}}}$ & 10.0 & 4.0 & 3.0 & 9.0 & - & ${4_{\text{p}}}$ & 10.0 & 3.0 & - & 9.0 & 1.3 & ${4_{\text{s}}}$ & 10.0 & 4.5 & 3.5 & 9.0 & - \\
    & & & & & & & ${4_{\text{p}}}$ & 10.0 & 3.0 & - & 9.0 & 1.38 & & & & & & & ${4_{\text{p}}}$ & 10.0 & 5.5 & - & 9.0 & 1.38
    \\                                            
\end{tabular} 
\end{ruledtabular}
\end{table*}

\renewcommand{\thetable}{A3}
\renewcommand{\tablename}{TABLE}
\setcounter{table}{0}
\begin{table}[!h]
 \caption{Parameters associated to Al and C basis sets are given. ${\text{r}_{\text{1, 2 or 3}}}$ are cutoff radii (Bohr) of each ``zeta'' for the orbital (orb). ${\text{r}_{\ast}}$ is the internal radius of the soft confinement potential\cite{Junquera2001}(with ${\text{V}_{0} \text{= 40 Ry}}$) while ${\text{r}_{\text{Q}}}$ is the radius associated to the charge confinement potential\cite{Corsetti2013}.}\label{tbl:Al-C basis}
\begin{ruledtabular}
  \begin{tabular}{c|ccccc}
 & orb & ${\text{r}_{\text{1}}}$ & ${\text{r}_{\text{2}}}$ & ${\text{r}_{\ast}}$ & ${\text{r}_{\text{Q}}}$ \\ \hline 
Al Basis & ${3_{\text{s}}}$ & 10.0 & 4.0 & 8.0 & - \\
    & ${3_{\text{p}}}$ & 10.0 & 5.0 & 8.0 & - \\ 
    & ${3_{\text{d}}}$ & 10.0 & - & 8.0 & 2.5 0.0 \\ \hline
C Basis & ${2_{\text{s}}}$ & 10.0 & 3.0 & 8.0 & - \\
    & ${2_{\text{p}}}$ & 10.0 & 2.8 & 8.0 & - \\ 
    & ${3_{\text{d}}}$ & 10.0 & - & 8.0 & 9.6 0.7 \\    
\end{tabular} 
\end{ruledtabular}
\end{table}

\renewcommand{\thefigure}{A1}
\setcounter{figure}{0}
\begin{figure}
\begin{center}
  \includegraphics[width=0.5\textwidth]{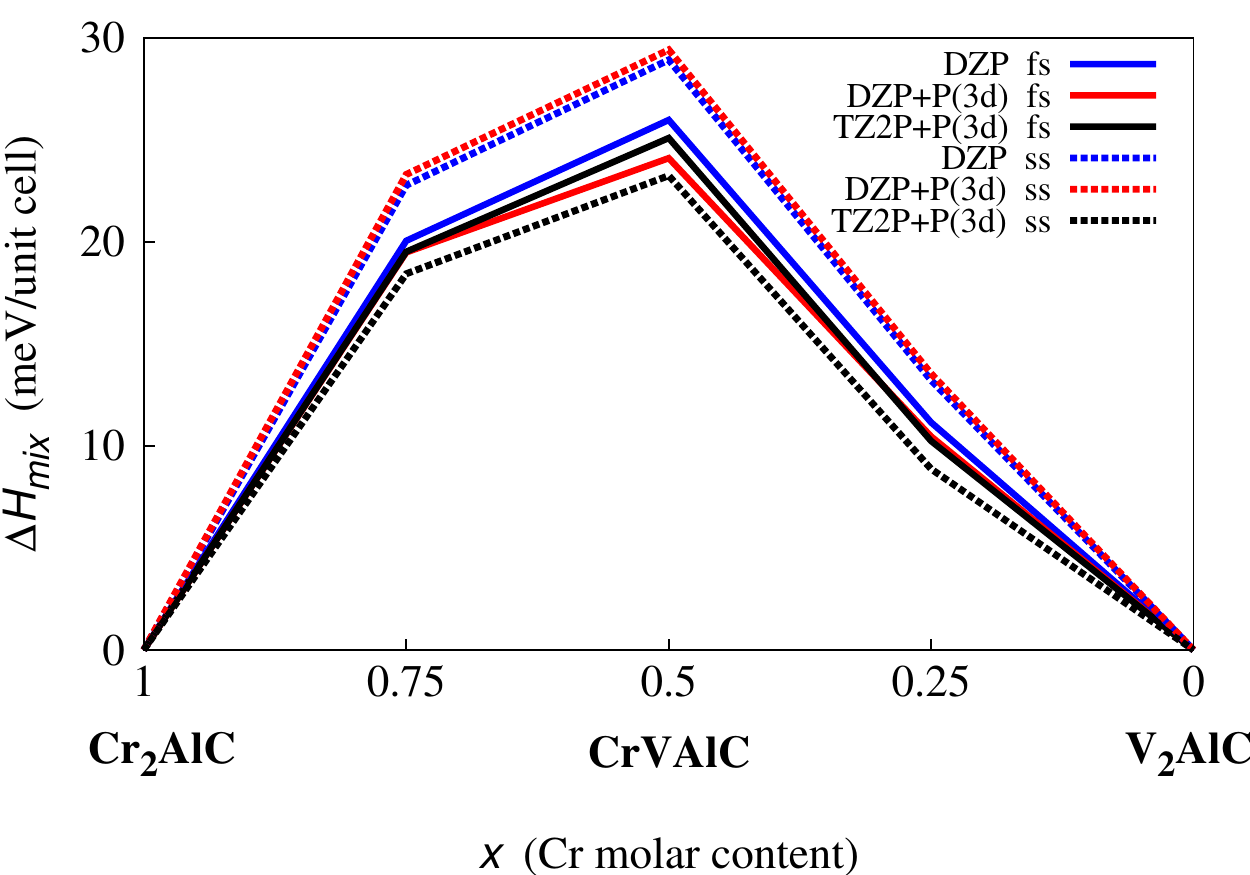}
\end{center}
\caption{Enthalpy of mixing ${\Delta H_{\text{mix}}}$ of the VVCrCr (NM) structure as a function of ${\,\,x}$, basis and pseudopotential.}\label{fig:basis-pseudos} 
\end{figure}

\section*{Appendix B}

The GGA${+U}$ calculations were performed using a value of ${U}$ = 1 eV for the Cr 3d electrons. The LDA${+U}$ projectors were generated as slightly-excited numerical atomic orbitals\cite{Riikonen2007} with a radius orbital of 2.3 Bohr along with a population and threshold tolerance of 0.0004 and 0.02 respectively. AFM spin arrangements for 2x1x1 chemically ordered structures and for VVCrCr\textbf{(}CrV\textbf{)}\textbf{(}VCr\textbf{)} are shown in Table B1 along with the corresponding electron populations with and without ${+U}$ corrections.  
\renewcommand{\thetable}{B1}
\renewcommand{\tablename}{TABLE}
\setcounter{table}{0}
\begin{table*}[t]
\caption{$2\times1\times1$ supercell structures for $\text{(}\text{Cr}_\text{0.5}\text{,}\text{V}_{\text{0.5}}\text{)}_{\text{2}}\text{AlC}$ denoted ${\text{VCrCrV}}$, ${\text{VCrVCr}}$, ${\text{VVCrCr}}$ and VVCrCr\textbf{(}CrV\textbf{)}\textbf{(}VCr\textbf{)}. The vertical direction points along the c-axis. The table gives the Mulliken spin population difference\cite{Mulliken1955} ${{n_{i}}_{\uparrow}-{n_{i}}_{\downarrow}}$ for atom ${i}$ and is listed in the same order along the c-axis as shown in the structures.These values were obtained for AFM spin arrangements and for the same spin arrangements with ${+U}$ corrections on the Cr atoms. Even though the ${+U}$ corrections did not affect the AFM arrangements for chemically ordered structures, for VVCrCr\textbf{(}CrV\textbf{)}\textbf{(}VCr\textbf{)} the spin arrangement changed to ferrimagnetic with a net magnetic moment given by 0.46 electrons.}\label{tbl:schematic-211-unit-cell} 
\begin{ruledtabular}
  \begin{tabular}{ccccc}
  &    
\minipage{0.20\textwidth}
  \centering	
  \includegraphics[width=\linewidth]{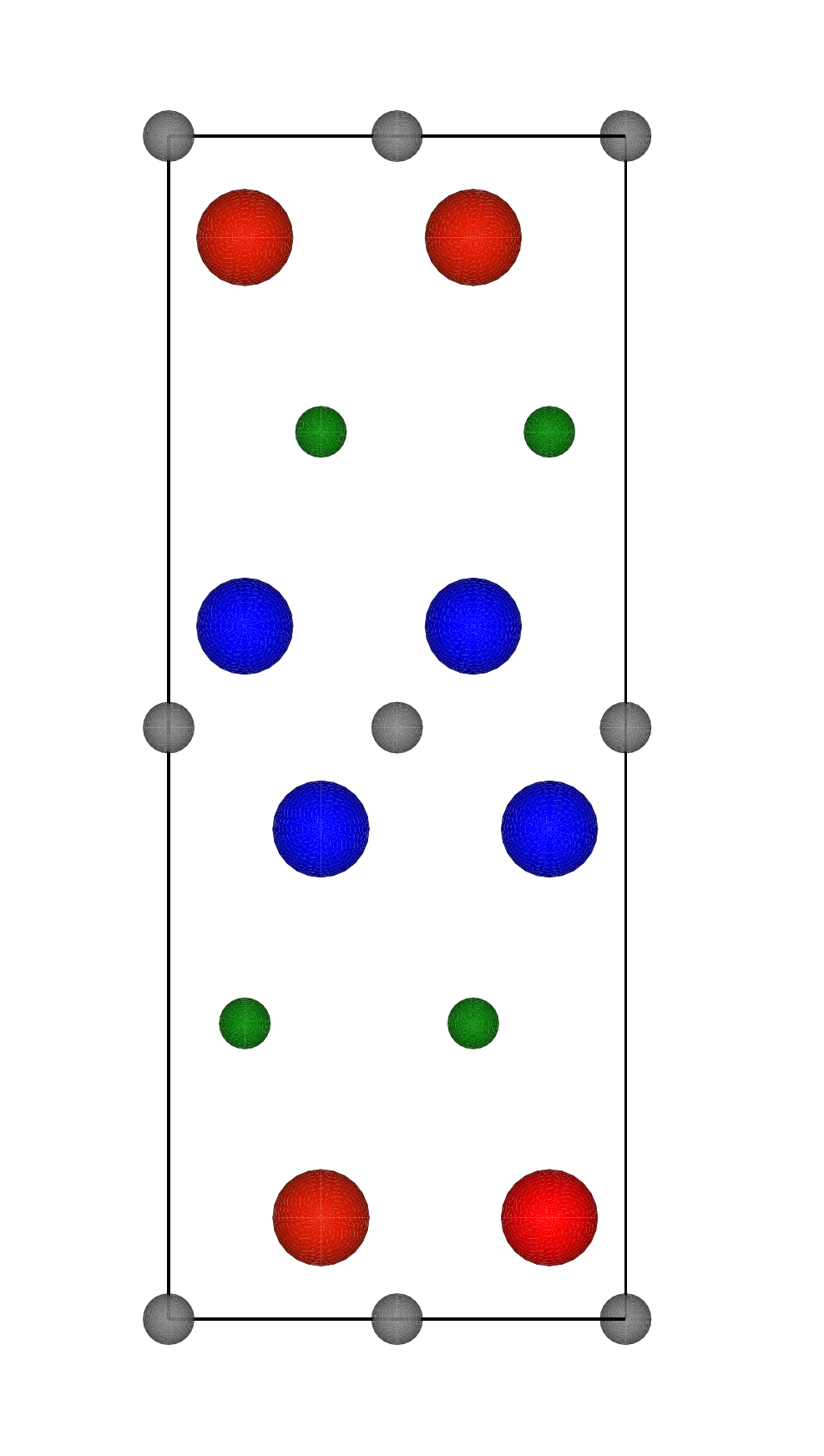}
  ${\text{2x1x1}\,\,\,\text{VCrCrV}}$
  \vspace{3mm}
\endminipage
&
\minipage{0.20\textwidth}
  \centering
  \includegraphics[width=\linewidth]{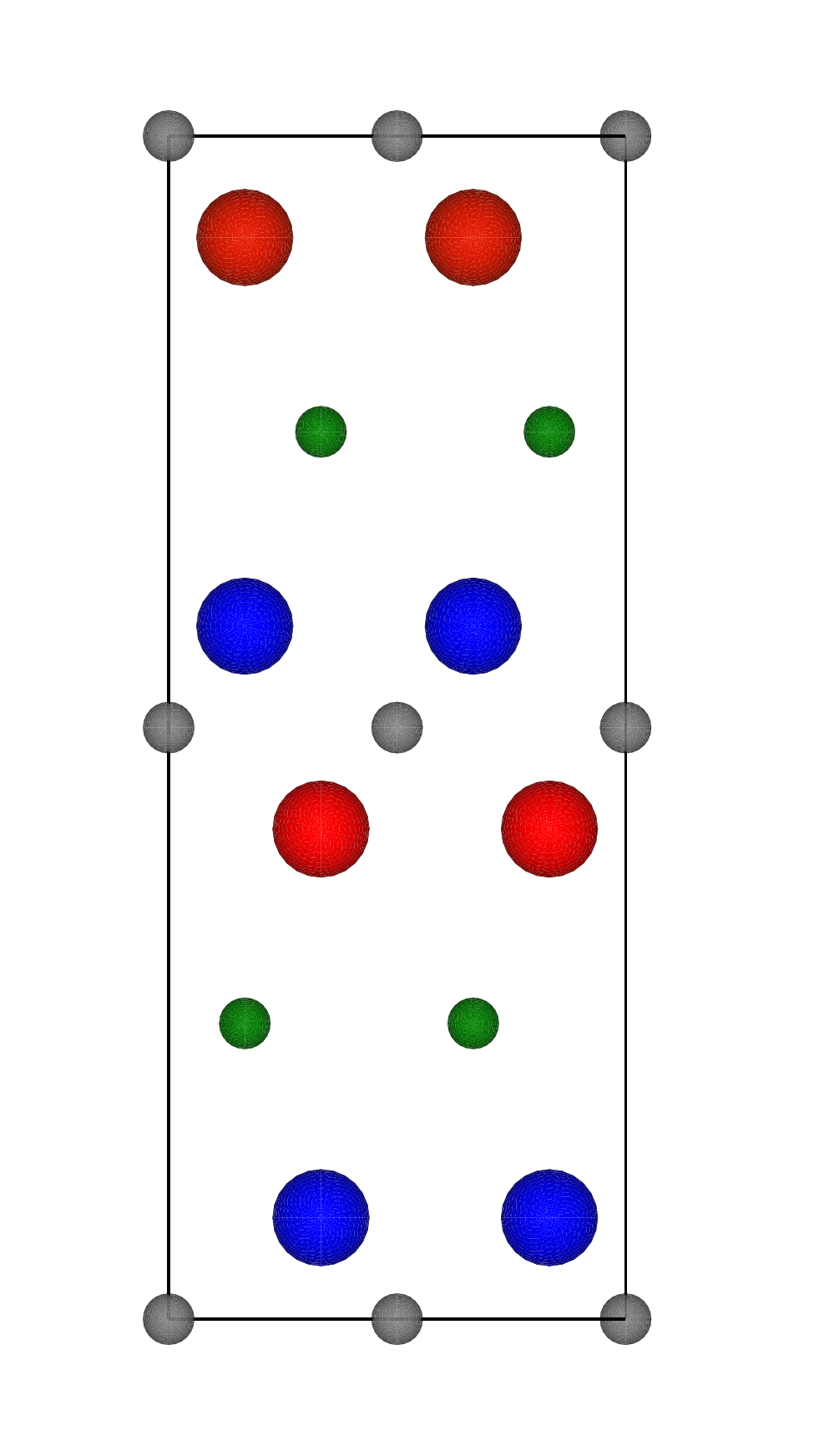}
  ${\text{2x1x1}\,\,\,\text{VCrVCr}}$ 
  \vspace{3mm}
\endminipage
&
\minipage{0.20\textwidth}
  \centering 
  \includegraphics[width=\linewidth]{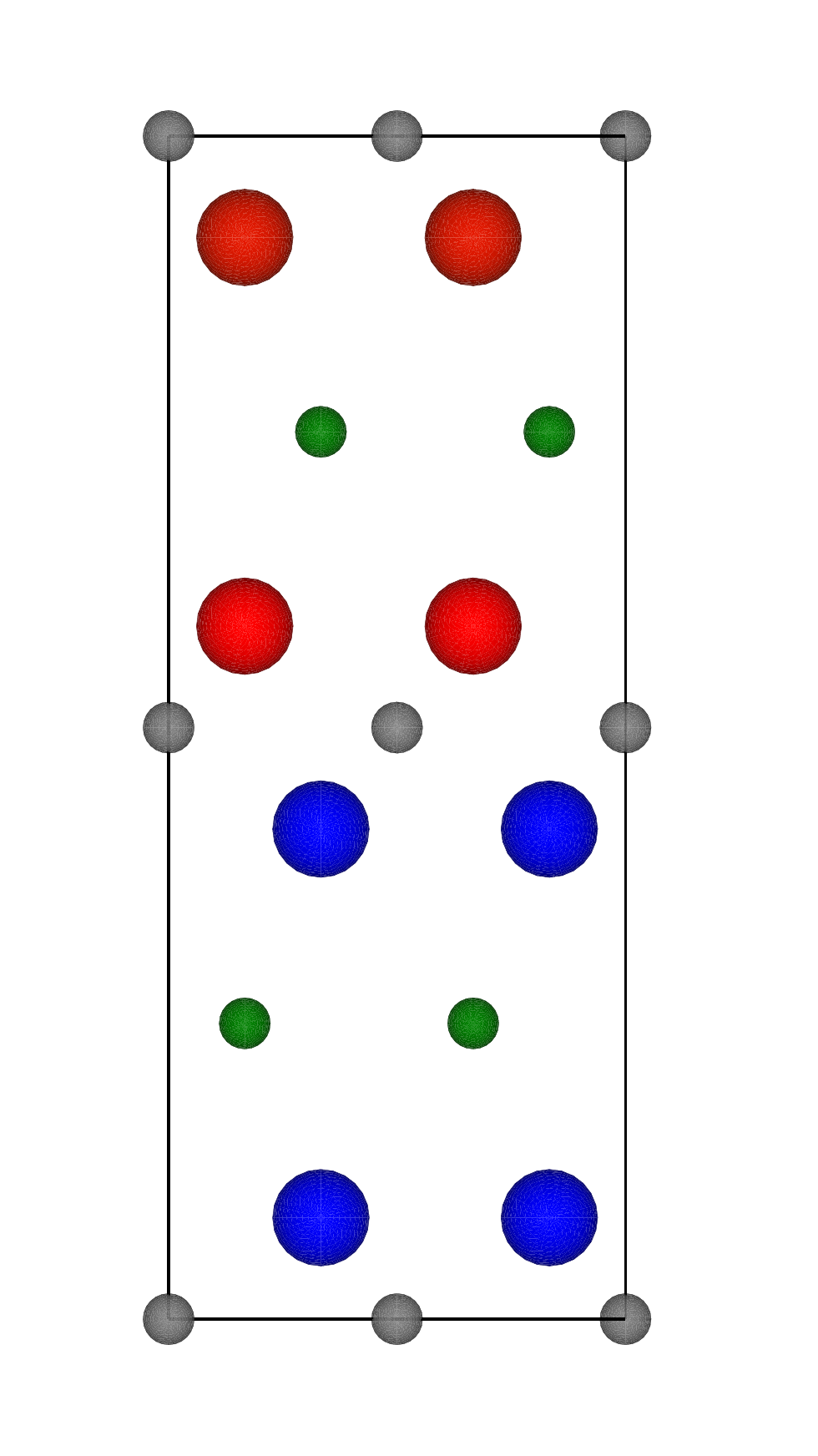}
  ${\text{2x1x1}\,\,\,\text{VVCrCr}}$
  \vspace{3mm}
\endminipage
&
\minipage{0.20\textwidth}
  \centering
  \includegraphics[width=\linewidth]{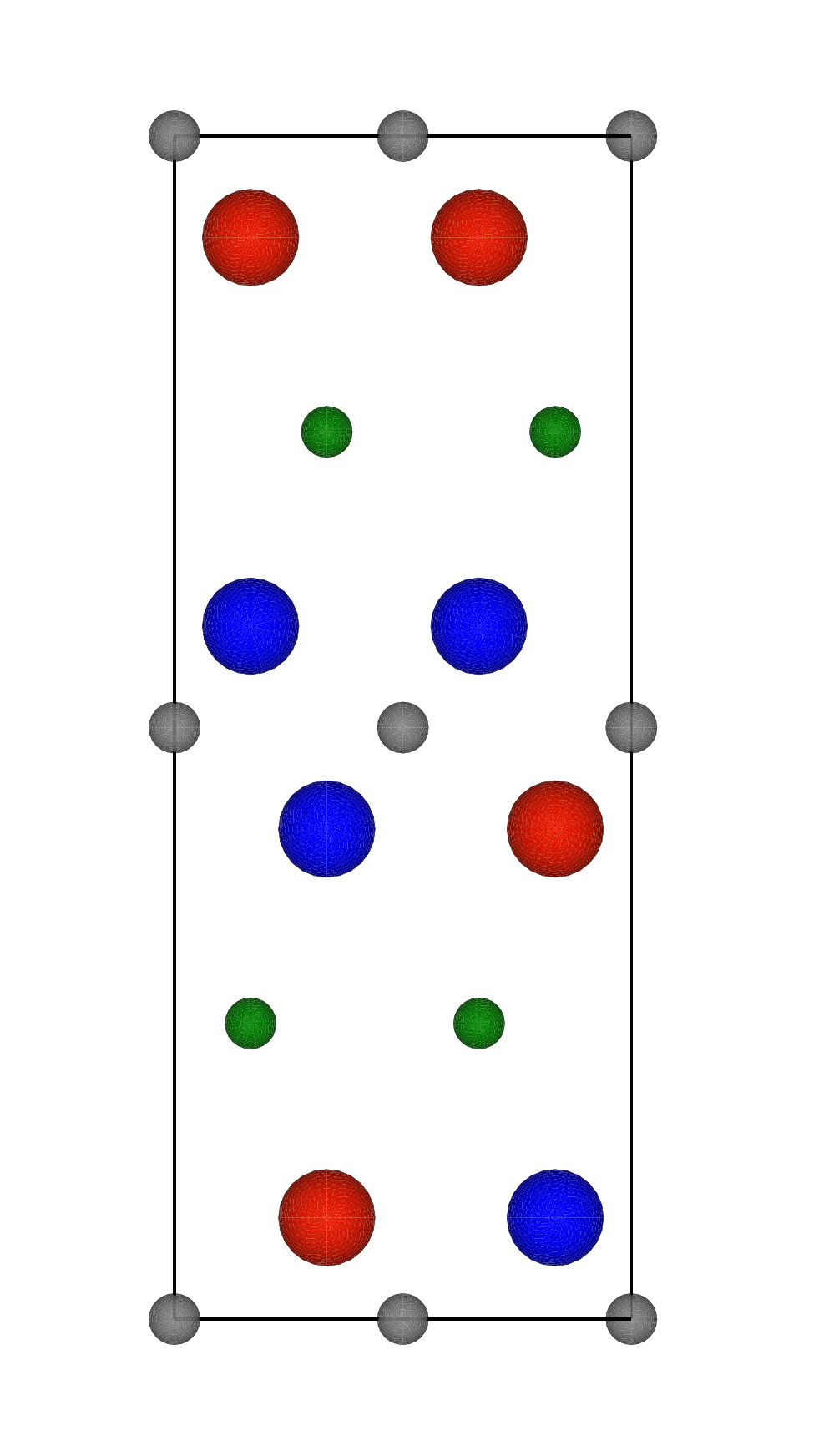}
  ${\text{VVCrCr}\textbf{(}\text{CrV}\textbf{)}\textbf{(}\text{VCr}\textbf{)}}$
  \vspace{3mm}
\endminipage \\
\hline \\
AFM & -0.10 {\color{red}V}${\,\,\,}$ +0.10 {\color{red}V}  
    & +0.01 {\color{red}V}${\,\,\,}$ -0.01 {\color{red}V}
    & -0.02 {\color{red}V}${\,\,\,}$ +0.02 {\color{red}V}
    & +0.04 {\color{red}V}${\,\,\,}$ -0.16 {\color{red}V}\\
    & -0.88 {\color{blue}Cr}${\,\,\,}$ +0.88 {\color{blue}Cr}
    & +0.08 {\color{blue}Cr}${\,\,\,}$ -0.08 {\color{blue}Cr}
    & -0.02 {\color{red}V}${\,\,\,}$ +0.02 {\color{red}V}
    & +0.52 {\color{blue}Cr}${\,\,\,}$ -0.67 {\color{blue}Cr}\\
    & +0.88 {\color{blue}Cr}${\,\,\,}$ -0.88 {\color{blue}Cr}
    & +0.01 {\color{red}V}${\,\,\,}$ -0.01 {\color{red}V}
    & +1.39 {\color{blue}Cr}${\,\,\,}$ -1.39 {\color{blue}Cr}
    & -1.26 {\color{blue}Cr}${\,\,\,}$ +0.19 {\color{red}V}\\
    & +0.10 {\color{red}V}${\,\,\,}$ -0.10 {\color{red}V} 
    & +0.08 {\color{blue}Cr}${\,\,\,}$ -0.08 {\color{blue}Cr}
    & +1.39 {\color{blue}Cr}${\,\,\,}$ -1.39 {\color{blue}Cr}
    & -0.24 {\color{red}V}${\,\,\,}$ +0.77 {\color{blue}Cr}\\ \\ \hline \\
AFM+U & -0.24 {\color{red}V}${\,\,\,}$ +0.24 {\color{red}V}  
    & +0.30 {\color{red}V}${\,\,\,}$ -0.30 {\color{red}V}
    & -0.03 {\color{red}V}${\,\,\,}$ +0.03 {\color{red}V}
    & +0.36 {\color{red}V}${\,\,\,}$ -0.32 {\color{red}V}\\
    & -2.19 {\color{blue}Cr}${\,\,\,}$ +2.19 {\color{blue}Cr}
    & +2.51 {\color{blue}Cr}${\,\,\,}$ -2.51 {\color{blue}Cr}
    & -0.03 {\color{red}V}${\,\,\,}$ +0.03 {\color{red}V}
    & +2.12 {\color{blue}Cr}${\,\,\,}$ -2.46 {\color{blue}Cr}\\
    & +2.19 {\color{blue}Cr}${\,\,\,}$ -2.19 {\color{blue}Cr}
    & +0.30 {\color{red}V}${\,\,\,}$ -0.30 {\color{red}V}
    & +2.36 {\color{blue}Cr}${\,\,\,}$ -2.36 {\color{blue}Cr}
    & -2.25 {\color{blue}Cr}${\,\,\,}$ +0.31 {\color{red}V}\\
    & +0.24 {\color{red}V}${\,\,\,}$ -0.24 {\color{red}V} 
    & +2.51 {\color{blue}Cr}${\,\,\,}$ -2.51 {\color{blue}Cr}
    & +2.36 {\color{blue}Cr}${\,\,\,}$ -2.36 {\color{blue}Cr}
    & -0.43 {\color{red}V}${\,\,\,}$ +2.21 {\color{blue}Cr}\\ \\ \hline 
&

&
\minipage{0.025\textwidth}
  \centering	
  \includegraphics[width=\linewidth]{Cr.pdf}
\endminipage
\,\,\,\,
\minipage{0.025\textwidth}
  \centering	
  \includegraphics[width=\linewidth]{V.pdf}
\endminipage 
\,\,\,\,
\minipage{0.0145\textwidth}
  \centering	
  \includegraphics[width=\linewidth]{Al.pdf}    
\endminipage
\,\,\,\,
\minipage{0.0135\textwidth}
  \centering	
  \includegraphics[width=\linewidth]{C.pdf} 
\endminipage
& 

& \\ 

& & Cr\,\,\,\,\,\,\,\,\,V\,\,\,\,\,\,\,Al\,\,\,\,\,C & & 
\end{tabular}
\end{ruledtabular}  
\end{table*}


\end{document}